\documentclass{article}
\usepackage[utf8]{inputenc}
\usepackage{amsmath,color,amssymb,graphicx,hyperref,ulem,multirow,pinlabel,hhline}
\usepackage[paper=letterpaper,margin=1.378in]{geometry}
\usepackage[authoryear]{natbib}
\usepackage{tikz}
\usetikzlibrary{positioning}
\definecolor {processblue}{cmyk}{0.96,0,0,0}
\newcommand{\ud}{\mathrm{d}}
    \newenvironment{dedication}
        {\vspace{3mm}\begin{quotation}\begin{center}\begin{em}}
        {\par\end{em}\end{center}\end{quotation}\vspace{1mm}}

\definecolor{darkblue}{rgb}{0.0,0.0,0.3}
\hypersetup{colorlinks,breaklinks,
        linkcolor=darkblue,urlcolor=darkblue,
        anchorcolor=darkblue,citecolor=darkblue}
\usepackage{mathpazo}
\title{ A Population Dynamics Approach to the Distribution of Space Debris in Low Earth Orbit }
\author{John Jurkiewicz\thanks{Department of Mathematical Sciences, University of Wisconsin - Milwaukee, Milwaukee, WI 53201, USA; Corresponding author: jurkiew4@uwm.edu} \and  Peter Hinow$^*$}
\date{\today}

\begin{document}
\maketitle

\begin{dedication}
Dedicated to the memory of Professor Ching-Shan Chou.
\end{dedication}
\begin{sloppypar}
\begin{abstract}
The presence of debris in Earth's orbit poses a significant risk to human activity in outer space. This debris population continues to grow due to ground launches, loss of external parts from space ships, and uncontrollable collisions between objects. A computationally feasible continuum model for the growth of the debris population and its spatial distribution is therefore critical.  Here we propose a diffusion-collision model for the evolution of debris density in Low-Earth Orbit (LEO) and its dependence on ground-launch policy. We  parametrize this model and test it against data from publicly available object catalogs to examine timescales for uncontrolled growth. Finally, we consider sensible launch policies and cleanup strategies and how they reduce the future risk of collisions with active satellites or space ships. 
\end{abstract}


\section{Introduction}\label{intro}

Since the beginning of human space exploration in 1957, the regions of outer space in proximity to Earth have been inhabited by artificial objects. While the population of near-Earth space in the period immediately following this time was comprised of active man-made bodies, the abandonment of satellites, inter-object collisions or unitary explosions have led to an ever-increasing presence of so-called space debris - inactive, non-controllable entities evolving according to Keplerian laws. This filling population risks runaway collision of existing debris objects and poses an increasing danger not only to safe operation of active spacecraft, but to any object existing in near-Earth space.  Indeed,  in 2022 a discarded rocket stage collided with the Moon causing a noticeable crater \citep{Moon}. Investigation of the activity and evolution of this debris environment is therefore an important direction of research.

The theory of artificial satellites has a rich history beginning  in  the early 1950s. Traditionally, near-Earth space is separated into three canonical altitude regions, the Low Earth Orbit (LEO,  altitude $\le$ 2,000  km), the Medium Earth Orbit (MEO, 2,000 - 36,000 km), and the Geosynchronous Earth Orbit (GEO, $\ge$ 36,000 km). In each of these three broad regions, the dominant forces and characteristics of existing objects differ substantially, and therefore each region requires different scientific treatment \citep{Rossi2005}. Of these three the GEO region is the most resistant to analysis since technological limitations mean that observations are limited to objects above a size of about 30 cm. Further, debris in the GEO region is not as impacted by the gravitational effects of the Earth and therefore does not exhibit decaying orbital patterns. Instead the gravitational effects of sun and moon as well as the solar radiation are the dominant influences on the behavior of objects in this region. The orbits of objects in this region are by necessity rather rather static and predictable, as active satellites must maintain a precise geostationary position above a point on Earth's surface. For this reason,  and the low density of cataloged objects in GEO (541 active satellites as of February 2022 \citep{GeoSat})  the GEO region density is generally considered to be stationary.

\begin{figure}[ht]
\centering
$\vcenter{\hbox{\includegraphics[width=70mm]{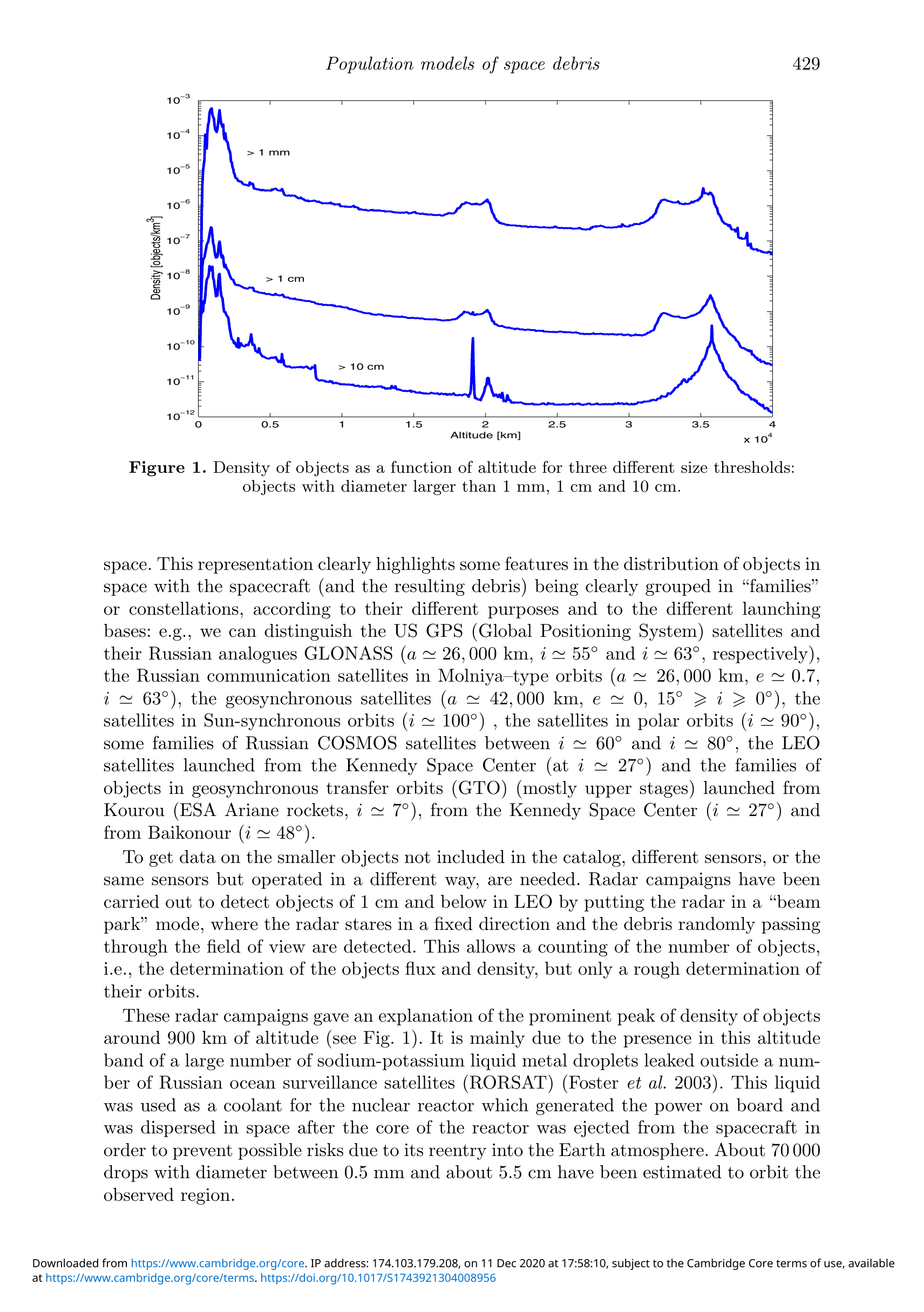}  }}$
$\vcenter{\hbox{\includegraphics[width=70mm]{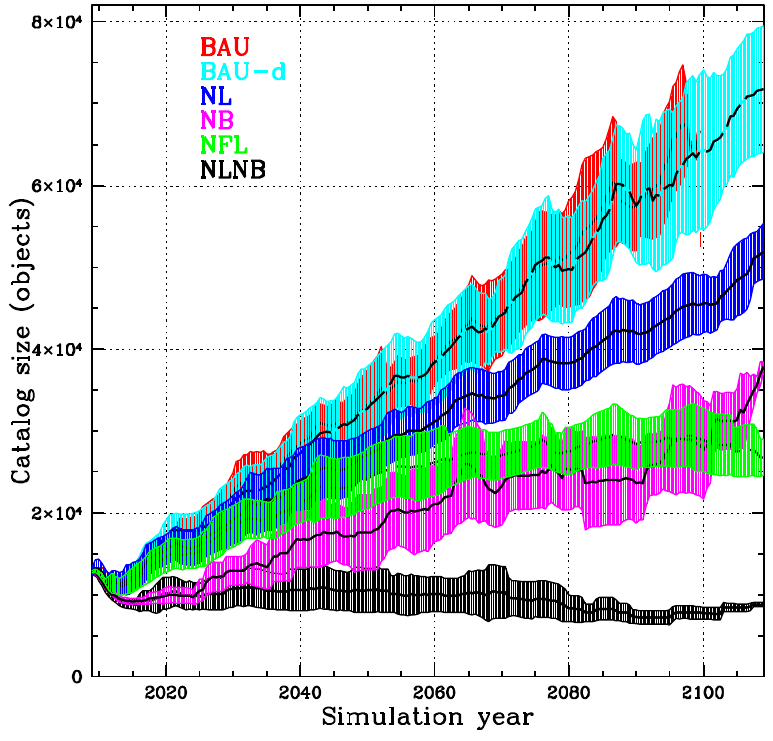}  }}$
\caption{\textbf{(Left)} Distribution of space debris of various sizes as a function of altitude. The LEO (left), MEO (center) and GEO (right) peaks are clearly visible. From \citep{Rossi2005}; reproduced with permission from Cambridge University Press.   \textbf{(Right)} Simulated growth of the catalog size of known objects under various mitigation strategies up to the year 2110. The ``business as usual'' (BAU, top funnel) is the most pessimistic scenario. From \citep{Nikolaev2012}; reproduced with permission of the Advanced Maui Optical and Space Surveillance Technologies Conference (\href{www.amostech.com}{\texttt{www.amostech.com}}). 
%\textcolor{red}{Slopes for comparison with our results: common starting point (approximately) 1.5$\cdot$10$^4$ in 2020.  BAU: reaches 4.5$\cdot$ 10$^4$ in 2080; 666 y$^{-1}$;  NL: reaches 3.8$\cdot$ 10$^4$ in 2080; 383 y$^{-1}$; NB (starts at 10$^4$),  reaches 2.5$\cdot$ 10$^4$ in 2080; 250 y$^{-1}$}. 
Note that this is the ``catalog size'', and it does not include spatial localization. }\label{Rossi1}
\end{figure}

The region most important to human space activity is LEO, due to the abundance of human activity. High-velocity collisions ($\approx 10$ km s$^{-1}$) have occasionally resulted in the destruction of active satellites such as the Iridium 33-Kosmos 2251 collision in 2009. Unitary satellite breakup or destruction is also a threat to safe operations. Examples are the aftermath of the intentional destruction of the Fengyun 1C weather satellite by a Chinese anti-satellite missile in 2007, or the destruction of Kosmos-1408 in an Russian anti-satellite weapon test in November 2021  \citep{Carnegie}. Each of these situations results in a net addition to the debris population that,  due to the density of extant object,  increases the probability of future debris interactions. As early as 1978, \citet{Kessler1978} proposed a cascade scenario in which collisions create new space debris that further endangers active objects, ultimately resulting in runaway evolution of the orbital object population. This is now known as the Kessler syndrome. A multitude of mathematical models and simulation software exists to predict the behavior of catalogued objects above a certain minimum size, e.g.~the EVOLVE \citep{Reynolds1995} and LEGEND \citep{Liou2004} model families. However, there are estimates of millions of unobserved objects in the 1 - 10 mm size range that are still capable of causing significant damage, and their undetected nature makes prediction difficult. 

At the level of tracking individual objects, the gold standard in analysis of artificial satellites has been the Simplified General Perturbations model (SGP, currently on SGP4) \citep{Hoots1980}. This model, based on the work of   \citet{Brouwer1959,Lyddane1963} assumes that ideal Keplerian orbits of satellites in an Earth-based reference system are perturbed by certain zonal harmonics that cause significant deviations from the ideal. Analytical solutions to the equations of motion for objects in such perturbed orbits allow observed state vectors of orbital objects to be propagated in time. This is an effective means of orbit determination for individual orbits over short timescales, and hence ideal for short-term policy and individual tracking. Nevertheless, this method   suffers from certain shortcomings. This model requires input data to be provided in the format of NORAD two-line element (TLE) sets, which is a very particular and obfuscating data format to encode and decode orbital state vectors. Secondly, this model is a single-particle model, which requires independent application to each cataloged object in orbit. Therefore, propagating the entire population of orbital debris potentially requires tens of thousands of separate applications. Thirdly, over long timescales the predicted orbit can vary significantly from reality \citep{Riesing2015}. Finally, the model itself does not directly include the presence of inter-object collisions or single-object explosions, which, though rare, cannot be neglected. 

There have been many attempts to incorporate collisions and explosions into applications of the SGP4 model, or similar. \cite{Nikolaev2012}, for example, utilizes a modeling approach, computing the trajectory of every known object of size $\geq$10 cm. In the event of a detected collision, well-known empirical distributions for number and size of particles created in collisions are used to update the catalog with new objects, whose trajectories are subsequently propagated along with the original objects as an enlarged cohort, to account for population growth. NASA's benchmark is the LEGEND model \citep{Liou2004,NASA}. The LEGEND approach is similar to that of \cite{Nikolaev2012}, but more all-encompassing as the entire space up to GEO is simulated in this model. Additionally, proprietary propagators are used in place of the standard SGP4. This has been the main tool used by NASA to study the near-earth environment since its completion in 2004. While very accurate, these methods still require propagation of thousands of individual orbits and sophisticated collision detection. Hence, an analytical population-dynamics model can possibly incorporate the key features of orbital evolution, collisions and explosions combined with significant computational savings.

Analytical models at the population level were first introduced by \citet{McInnes1993}, see also \citep{Zhang2019} for a more recent contribution. Although space debris is by definition non-biological material, the mathematical modeling approach to its behavior shares a surprising similarity to that of spatially structured biological populations that are subject to birth and death processes, diffusion, as well as inter-species interactions \citep{Okubo1980}. The model of \citet{McInnes1993} includes a convection term and a growth term due to binary collisions. In addition, space debris is subject  to the effect  of solar wind which is included as a diffusion term. In this work, we propose a spherically symmetric diffusion model with interaction term for space debris population evolution. We examine public databases of existing TLEs to initialize the model and set parameters for the subsequent numerical simulations by comparisons with  TLEs, the literature, and the LEGEND population predictions. The diffusion parameter varies with altitude, thereby incorporating the effects of both atmospheric drag, gravitational perturbations and solar wind. The use of a novel analytical model allows for significant computational savings and avoidance of numerical instability while capturing the important mechanisms of population growth and providing good  approximation to debris activity.  Moreover, we allow for explicit control at the level of launch policy and potential cleaning mechanisms. This provides the framework for testing of different strategies to mitigate the growth of the population density. Our work therefore has the potential to inform future space law and policy \citep{Gast2022}. 

\section{Introducing a spatial population model for LEO}\label{modeling}

\begin{figure}[ht]
\centering
$\vcenter{\hbox{\includegraphics[width=65mm]{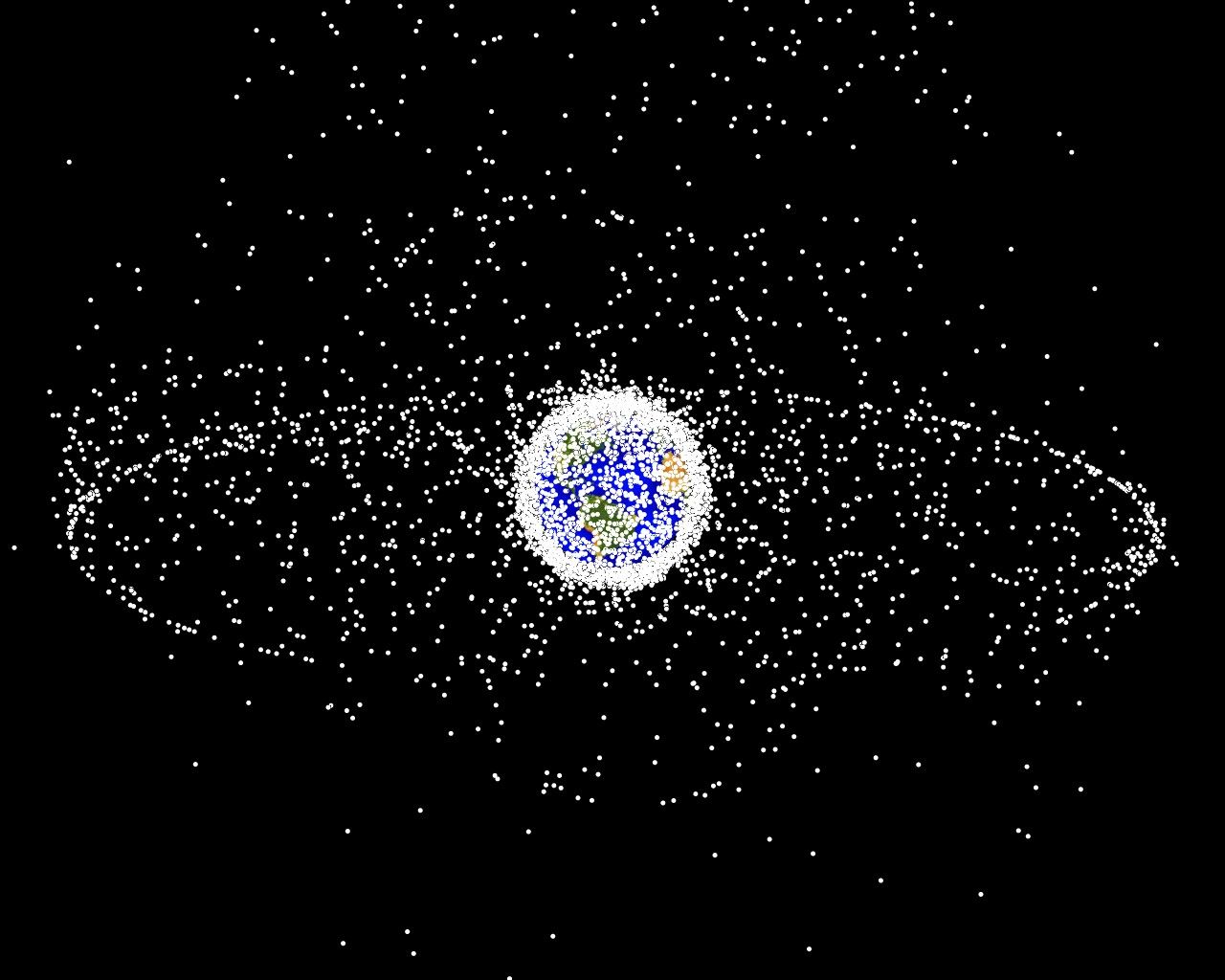}}}$  
$\vcenter{\hbox{\includegraphics[width=74mm]{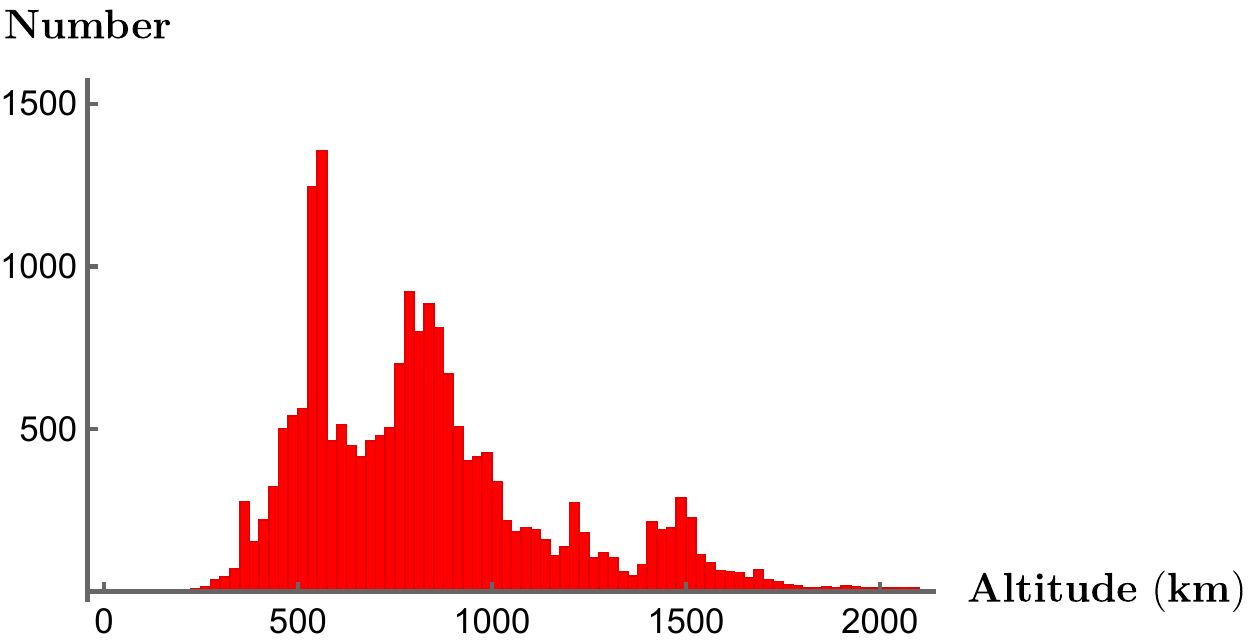}}}$  
\caption{\textbf{(Left)} A computer generated image of space debris as seen from high Earth orbit. Clearly visible are the LEO cloud and the GEO ring. From NASA's Orbital Debris Program Office; available at \href{https://orbitaldebris.jsc.nasa.gov/photo-gallery/}{\texttt{orbitaldebris.jsc.nasa.gov/photo-gallery}.} \textbf{(Right) }  Histogram snapshot of spatial density of orbital population in LEO as of March 22nd, 2022. Histogram generated by obtaining TLEs of all catalogued objects, and extracting altitudes by use of \textsf{pyorbital}. This is distinct from the density functions used in Equation \ref{mod1}. } \label{NASA1}
\end{figure}

\begin{figure}[ht]
    \centering
    \includegraphics[width=70mm]{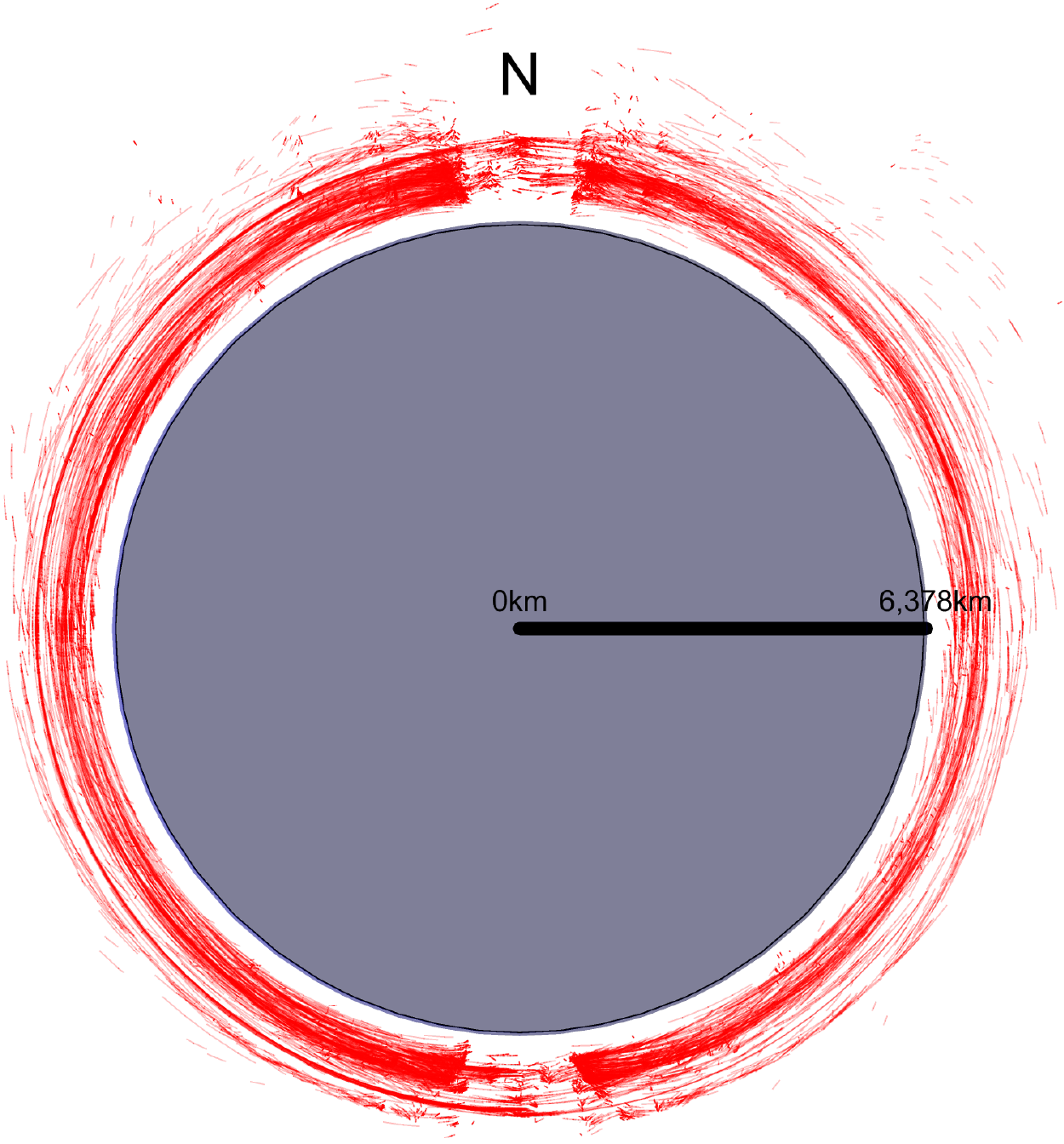}\hspace{5mm}
    \includegraphics[width=70mm]{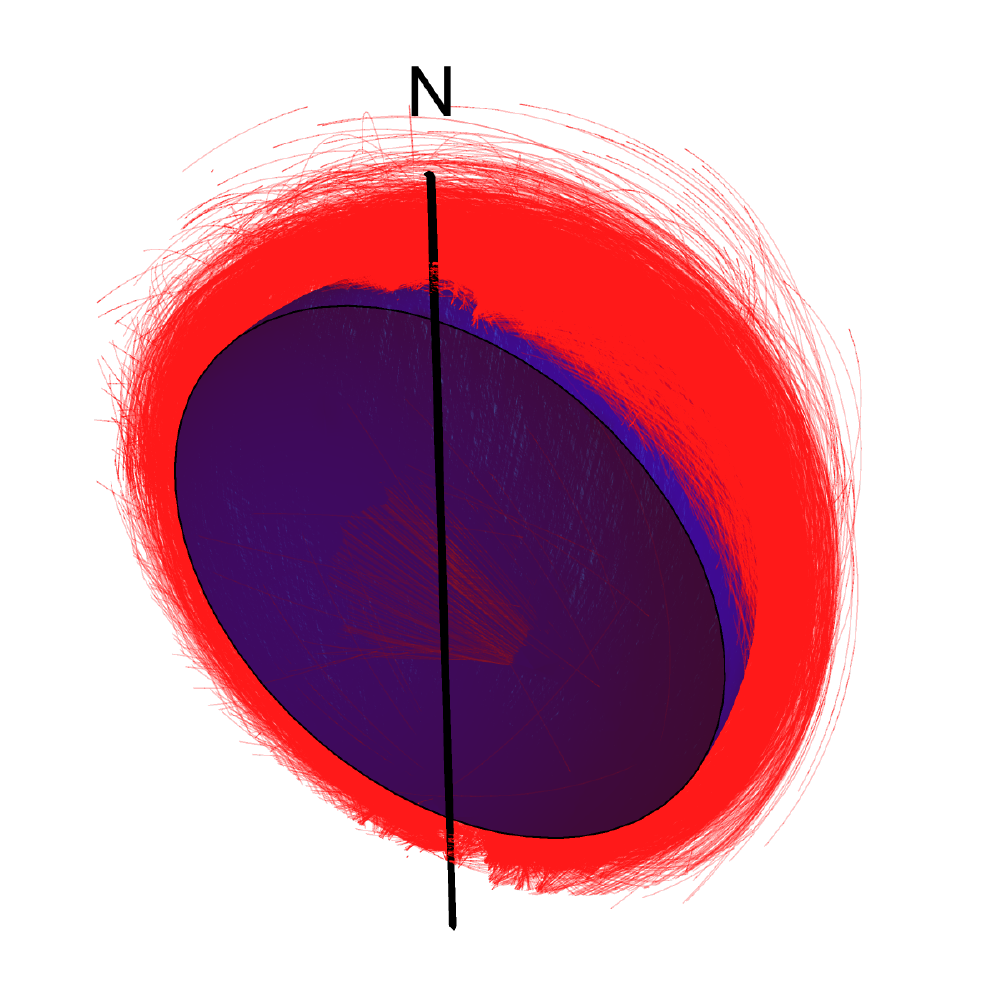}
    \caption{Cross-sectional views of the distribution of satellite orbits in LEO as of March 2nd, 2022. Aside from minor conical areas near the poles, orbital ellipses cover the volume of near-Earth space essentially uniformly. }
    \label{symmetry}
\end{figure}

\begin{figure}[ht]
\centering
$\vcenter{\hbox{\includegraphics[width=70mm]{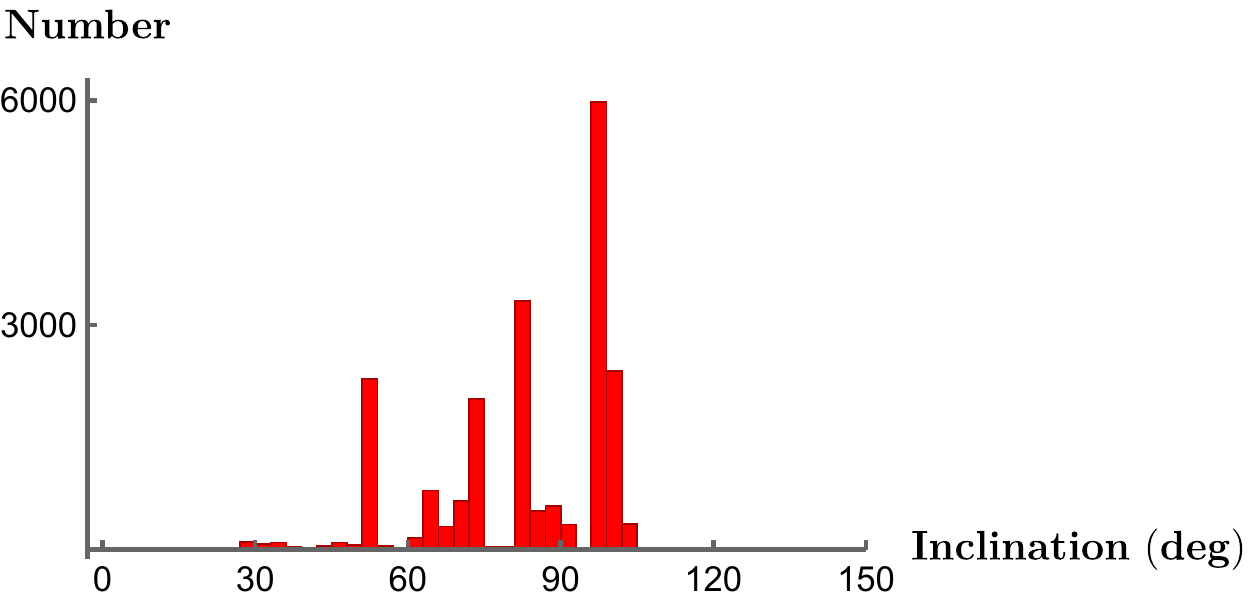}}}$
$\vcenter{\hbox{\includegraphics[width=70mm]{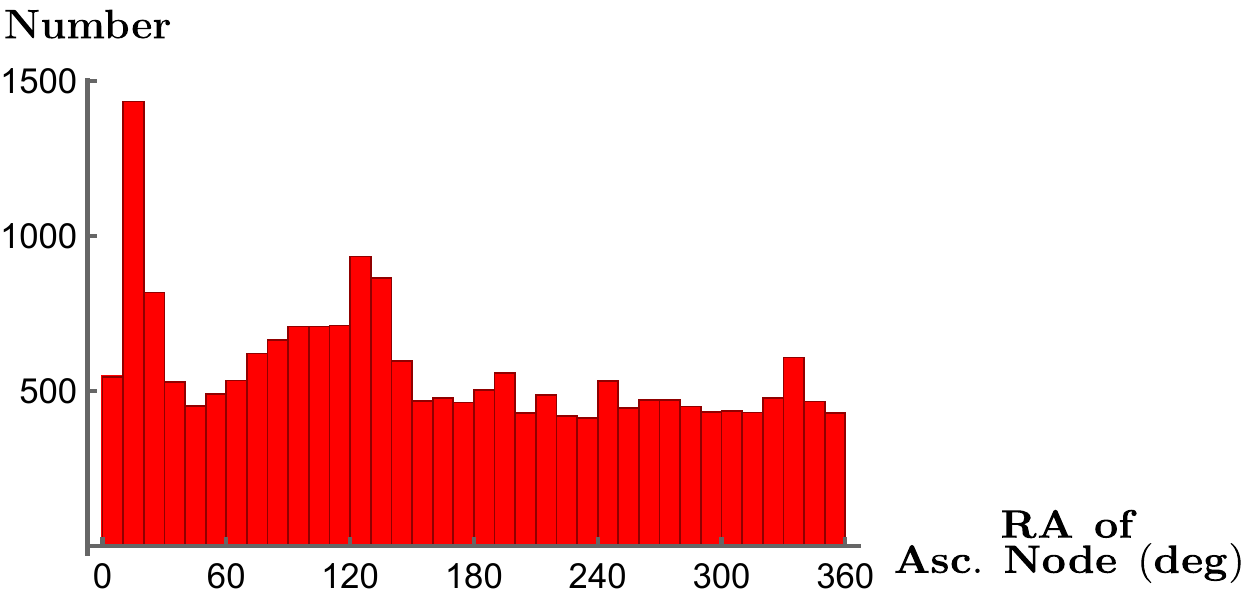}}}$

\caption{Histograms  of \textbf{(Left)}  orbital inclination and  \textbf{(Right)} right ascension (longitude) of ascending node (the longitude at which the orbit intersects Earth's orbital plane) as of March 2nd, 2022. These together show that spherical symmetry is a reasonable   assumption to model the dynamics of orbital debris.}\label{Histograms}
\end{figure}

We begin by stating the underlying assumptions on the structure of orbital debris. It is a natural assumption supported by empirical observations, as in Figures \ref{NASA1}-\ref{Histograms}, that the orbital population is spherically symmetric. Thus let $u(r,t)$ denote the density of orbital objects at time $t$ and a distance of $r$ from the center of the Earth.  Assuming the Earth to be perfectly spherical, the total population size is obtained by integration, \begin{equation}\label{total}
    U(t) = 4\pi \int_{r_E+200}^{r_E+2000} u(r,t)r^2\,\ud r,
\end{equation}
where $r_E=6378$ km is the equatorial radius of the Earth. 

LEO space is affected by several factors, namely gravitational effects from the Earth-Moon-Sun system, solar wind, and, in lower regions, atmospheric friction. To account for these factors on the behavior of the LEO population, we postulate that the spatial density $u(r,t)$ evolves in time according to a diffusion equation with source and collision term $\Phi$,
\begin{equation}\label{mod1}
    \frac{\partial}{\partial t} u(r,t)= \frac{1}{r^2}  \frac{\partial}{\partial r} \left(D(r)
    r^2\frac{\partial}{\partial r} u(r,t)\right) +\Phi(u,r).
\end{equation}
The diffusion term in this equation represents the combined effects of atmospheric drag, a  multi-body gravitational system and solar wind. Solar wind is assumed to cause a diffusive behavior on average, due to the periodic nature of orbits. In order to capture the altitude dependence of these perturbations, we select a piecewise radially dependent diffusion parameter  
\begin{equation}\label{diff}
D(r) = \left\{
	\begin{array}{cl}
		\alpha \exp(-\lambda (r-r_E)), & r<r_E+1000 \\
		\xi, & r\geq r_E+1000 \\ 
	\end{array}\right..
\end{equation}
Here $\alpha$ and $\xi$ are setting the diffusion rate in the near and the far field, respectively, and $\lambda$ sets the spatial scale. The choice of these values is discussed in detail in Section \ref{param}. The functional form is inspired by the presence of atmospheric drag which decreases exponentially with altitude. Above a certain altitude only the solar wind effects the trajectories. In accordance with the observations of \citep{Lemaitre2013}, we choose a critical altitude of $1000$ km as atmospheric drag works on a timescale of 10$^3$-10$^4$ years at this altitude.

The term $\Phi$  represents a ``birth and death process'' for the debris population. Debris is created from two sources namely collisions between present objects and deposition of new objects by ground launch.  We set 
\begin{equation}\label{birth}
 \Phi(u,r)=C(u)+\Delta(r,t)
\end{equation}
where $C(u)$ is the rate of object creation due to binary collisions between orbital objects, and $\Delta(r,t)$ is the rate of deposition of objects into orbit. We are at the moment ignoring explosions as a source but leave the possibility that these be incorporated as stochastic impulses in the future.  In \cite{McInnes1993}, a quadratic collision term of Smoluchowski type is proposed for binary collisions of space debris subject to certain averaging assumptions,
\begin{equation}\label{collision}
C(u)=\frac{\partial u(r,t)}{\partial t}|_{collision}=\frac{1}{\sqrt{2}}\beta\gamma\sqrt{\frac{GM}{r+r_E}}u^2(r,t) 
\end{equation}
where $G$ is the gravitational constant, $M$ is the mass of Earth,   $\beta$ is the average number of new particles created in a binary collision, and $ \gamma$ is the mean cross-sectional area of space debris. We assume that the deposition rate is separable, that is  
\begin{equation}\label{deposfct} 
    \Delta(r,t) = R(r)T(t).
\end{equation}
The idea is that there are ``natural'' locations for deposition of  working objects and that merely the amount varies over time. The functional form is allowed to vary when we consider different strategies to mitigate debris growth, but this separable form is the standard for business-as-usual computations.

The model is completed by a homogeneous Dirichlet boundary condition at the lowest altitude as objects that enter the denser layers of the atmosphere burn up (although occasionally large pieces of space debris have reached the surface of the Earth) and exit the population. At the upper altitude we impose a homogeneous Neumann boundary condition since there is no debris migrating from the MEO region. Thus we have 
\begin{equation}\label{bd}
    u(r_E+200,t)=0, \quad \frac{\partial}{\partial r}u(r_E+2000,t)=0
\end{equation}
for all $t\ge 0$.  Finally, we will discuss in Section \ref{param} the selection of the initial condition
\begin{equation}\label{in_cond}
    u(r,0)=u_0(r), \quad r\ge r_E+200.
\end{equation}

\section{Parametrization of the model }\label{param}
 
To validate the model and aid in setting certain parameter values, we compare our model's predictions to those of NASA's LEGEND model \citep{Liou2004} in the case of continued deposition with no mitigation \citep{NASA}. As a tool, LEGEND is presented to the public in a limited manner, and the available predictions are only in terms of the total population. Hence, comparison to our work can be done only by integrating the density at each time stamp and comparing the evolution of this total population to the predictions of LEGEND. We run our simulations with an artificially reduced launch rate to enable accurate comparisons to LEGEND's predictions obtained from \cite{Liou2011}. 

Free parameters and initial conditions are established using real-world orbital data in the TLE format obtained from Space-Track \citep{SpaceTrack}. As such, the data do not describe Keplerian orbits and are specialized for use with the simplified perturbations models \citep{Hoots1980}. This necessitates the use of specialized software to decode the TLE data into Keplerian elements that can be used to compute position vectors. We use the open-source Python distribution  \textsf{pyorbital} to extract the osculating orbital data  \href{https://pypi.org/project/pyorbital}{\texttt{(pypi.org/project/pyorbital)}}. The altitude at any given epoch is computed from the most recent TLE preceding the chosen epoch for each cataloged object to obtain the cumulative population function. This function is then differentiated and divided by $4\pi r^2$, as $r$ ranges from $r_e+200$ to $r_e+2000$, to obtain the initial condition $u_0(r)$ for Equation \eqref{in_cond}.  

\begin{figure}[ht]
    \centering
    \includegraphics[width=10cm]{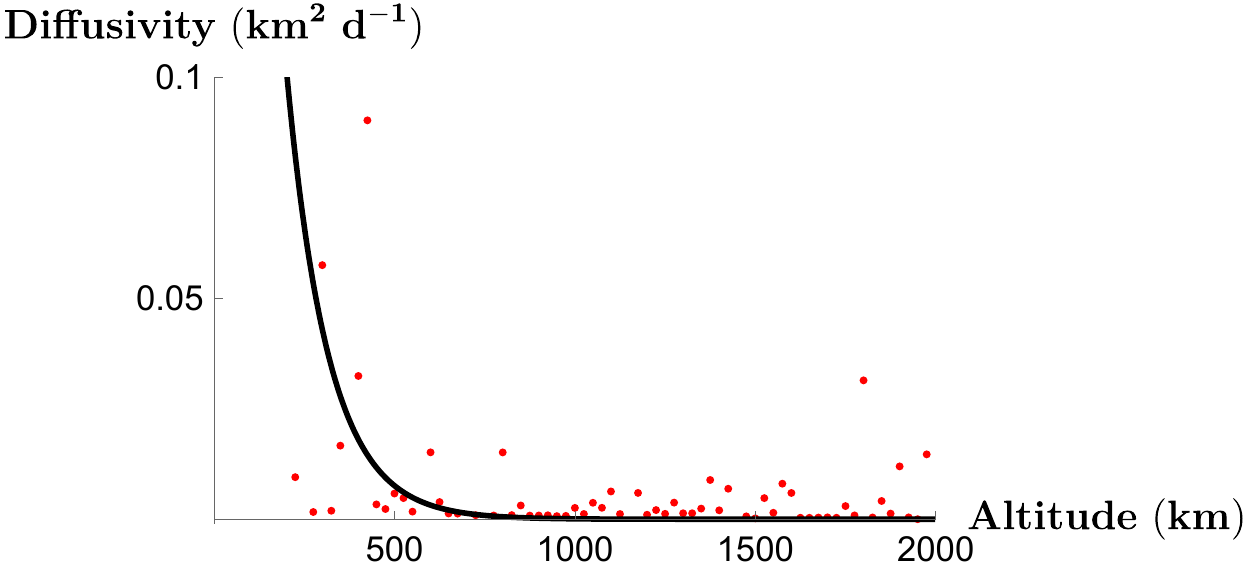}

    \caption{ Diffusivity function for Equation \ref{diff} for individual orbits at different altitudes.} 
    \label{Diffuse}
\end{figure}

To select the values of diffusion parameters $\alpha$ and $\xi$ for Equation \eqref{diff} we use variability present in the available TLE databases, as this variability is due to the effects of the same perturbing forces. In the same manner as \cite{Riesing2015}, we take the two most recent TLEs for every tracked object currently in LEO; the older TLE for each satellite is propagated to the epoch of the newer TLE, and the error in altitude (propagated to observed) is recorded, assuming that the more recent TLE gives the correct osculating value. We gather these errors into groups based on altitude and time difference in epoch and compute the standard deviation of each bin. For each group, we use the formula $\sigma=\sqrt{2Dt}$ to gather  diffusion rates. Finally at each altitude we take the average value of these diffusion rates to model the ``true" diffusivity. Choosing $\xi=10^{-4}$ km$^2$ d$^{-1}$ in keeping with the observations of \cite{Lemaitre2013} and the cataloged data, we fit the above piecewise exponential function to this scatter plot via nonlinear regression as implemented in Wolfram Mathematica, giving $\alpha=0.5783$ km$^2$ d$^{-1}$ and thus forcing $\lambda=0.0086$ km$^{-1}$. A plot of the diffusion rate as a function of the altitude is shown in Figure \ref{Diffuse}. 

The parameters $\beta$ and $\gamma$ in the collision term \eqref{collision} represent the number of new objects created in a binary collision and the average cross-sectional area of a piece of space debris, respectively. We set $\beta=2000$, taking the Iridium-Cosmos collision of 2009 as a representative \citep{Apollo}.   In practice, we set $\gamma$ to ensure a good fit in the $L^1$ and $L^\infty$-norms of our model to accepted standard predictions, namely those presented in LEGEND. This agreement requires the presence of a birth term as described in Section \ref{birth}. The resultant value of $\gamma$ is $9.98\cdot 10^{-8}$ km$^{-2}$, corresponding to a cross-sectional diameter of approximately $17$ cm. This is a sensible value, as the lower bound for observable diameter is $5-10$ cm. The parameters used in  Equations \eqref{mod1}-\eqref{in_cond} are  collected in Table \ref{modparam}.

\begin{table}[ht]
    \centering
    \begin{tabular}{|c||c|}
    \hline
      $\alpha$   & 0.5783 km$^2$ d$^{-1}$ \\ \hline
			$\xi$ &  $ 10^{-4}$ km$^2$ d$^{-1}$ \\ \hline
    $\lambda$     &  0.0086 km$^{-1}$\\ \hline%{|=||=|}
    $\beta$ & 2000 \\ \hline
    $\gamma$ &$9.98\times 10^{-8}$ km$^2$ \\
    \hline
    \end{tabular}
    \caption{Collected parameters for our model Equations \eqref{mod1}-\eqref{in_cond}.}
    \label{modparam}
\end{table}
\begin{figure}
    \centering
    \includegraphics[width=71mm]{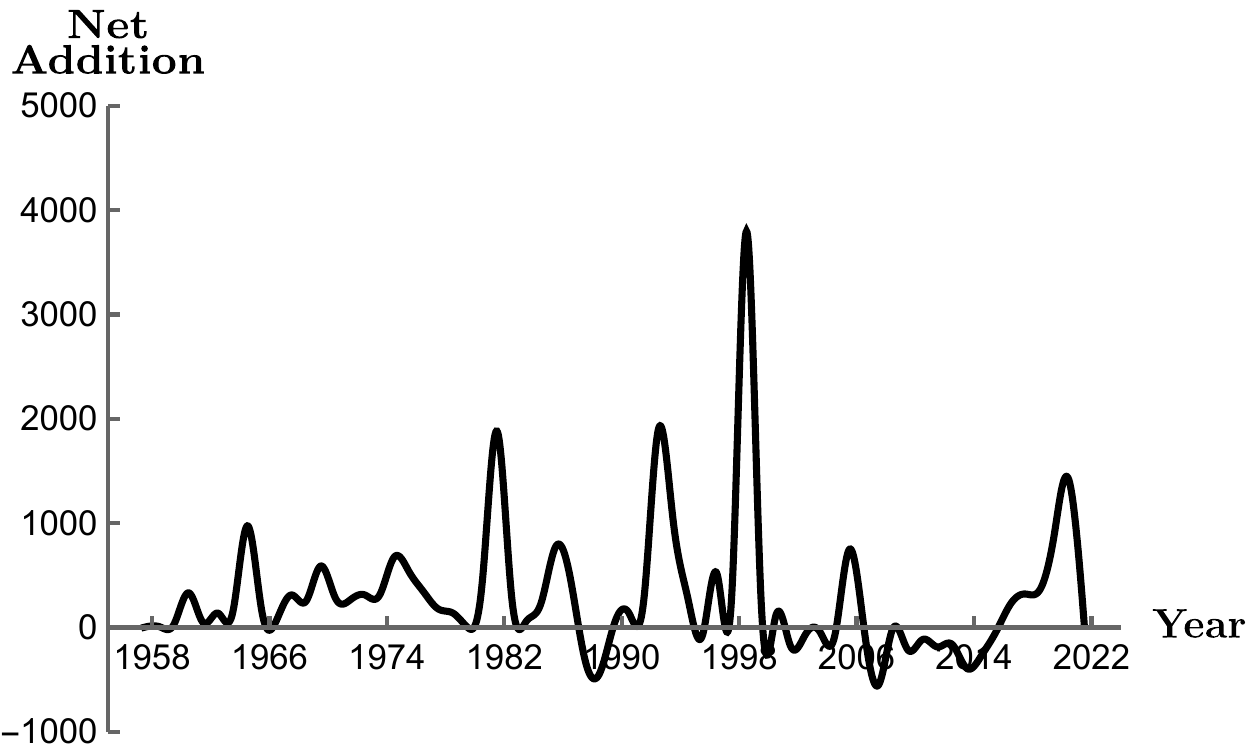}
     \includegraphics[width=71mm]{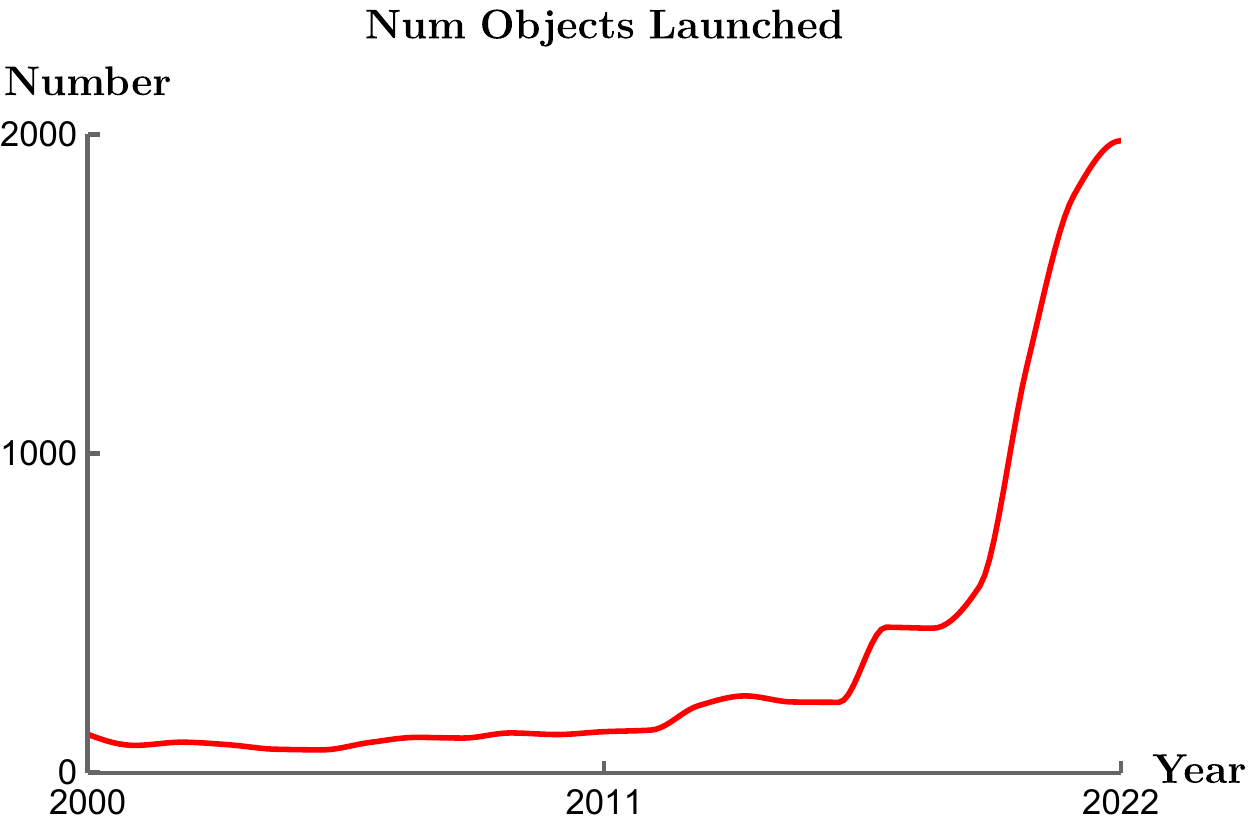}   
    \caption{\textbf{(Left)} Net addition of objects to the LEO population since 1957, exhibiting a period of approximately $T=5$ y.  \textbf{(Right)} Number of satellites launched into LEO each year from 2000 to 2021, maintained by the UN Office for Outer Space Affairs (accessible at \href{https://www.unoosa.org/oosa/osoindex/search-ng.jspx}{\texttt{https://www.unoosa.org/oosa/osoindex/search-ng.jspx}}). Note the increase in launches in 2019, the year Starlink was first launched \citep{Starlink}.}\label{DepositionRate}
\end{figure}

The space and time components of the deposition function from Equation \eqref{deposfct} are possibly the most difficult objects to estimate for our model. First we begin with 
\begin{equation*}
 R(r) = \sum_{k=1}^n a_k \exp\left(-\frac{(r-r_k)^2}{\sigma_k^2}\right),
\end{equation*}
where the $r_k$ are the typical orbit locations and the  $\sigma_k$ are the widths of those regions. This is motivated by inspection of the densities from 2014 through 2022; there appear to be certain ``natural" altitudes at which most objects are deposited. Our choice of values was made to give $R(r)$ a total mass of 1, and is given in Table \ref{param_R}. 

To set the $T(t)$ factor, we use the catalog of total number of objects launched for the last several years, maintained by the UN Office for Outer Space Affairs (accessible at \href{https://www.unoosa.org/oosa/osoindex/search-ng.jspx}{\texttt{https://www.unoosa.org/oosa/osoindex/search-ng.jspx}}). We interpolate this list of values via cubic spline interpolation, giving the variable launch rate for the years from 2000 to 2022. For years prior to 2022, this interpolation is the time-varying factor $T(t)$. Additionally, we make the simplifying assumption of uniform deposition over the year. Hence, if $n$ objects are deposited during a year, the effective rate is $\frac{n}{365}$ per day. This choice results in an agreement between the propagated and observed densities from 2014 through 2020, see Section \ref{simulation}.

\begin{table}[ht]
    \centering
    \begin{tabular}{|l||c|c|c|c|}
    \hline
      $k$   & 1 & 2 & 3 & 4  \\ \hline
	  $a_k$ (km$^{-3}$) &  $5.599\cdot 10^{-16}$  &  $1.39\cdot 10^{-11}$ &  $8.39\cdot10^{-12}$ & $1.39\cdot 10^{-11}$  \\ \hline
    $r_k$ (km) & 200 & 500 & 700 & 850 \\ \hline 
    $\sigma_k$ (km) & 7.07 & 20.09 & 100 & 9.98  \\
    \hline
    \end{tabular}
    \caption{Peak locations, widths and strengths of the deposition function $R(r)$.  }
    \label{param_R}
\end{table}

For propagation of current launch policies into the future, we note the essentially periodic behavior of net object addition into the LEO population, see Figure \ref{DepositionRate} left panel. Therefore, for future prediction, we multiply the current (2022) launch rate $I_{2022}$ by  \begin{equation*}
    P(t)=k_1+k_2\sin(2\pi f (t-t_0))
\end{equation*}
That is, we let the basic 2022 launch rate be perturbed periodically in future epochs. We take $f=0.2$ y$^{-1}$ based on Figure \ref{DepositionRate}, left panel. $k_1$ and $k_2$ are dimensionless free parameters that can be used to emulated different launch policies. In our analysis, we take $k_1=1$ and $k_2=5.02\cdot 10^{-4}$. Thus for future propagation our time-varying factor is 
\begin{equation}
    T(t)=I_{2022}P(t).
\end{equation}

\section{Simulation and comparison with data}\label{simulation}

We implement numerical simulations of  the model \eqref{mod1}-\eqref{in_cond} using the Crank-Nicolson method as the numerical solver. The spherical Laplacian in Equation \ref{mod1} is discretized by centered differences with an evenly-spaced sampling of $\Delta r=2.4$ km. In the time  direction $\Delta t$ ranges between $0.05$ d and $3.65$ d. Due to the small magnitude of the diffusivity, the ratio $\tfrac{D\Delta t}{(\Delta r)^2}\leq\tfrac{1}{2}$ for all of these values, and thus the method is numerically stable. All code used in this paper can be found at \href{https://github.com/jurkiew4/Space-Debris-Analytic-Model}{\texttt{https://github.com/jurkiew4/Space-Debris-Analytic-Model}}.

\begin{figure}[ht]
    \centering
    $\hbox{\includegraphics[width=71mm]{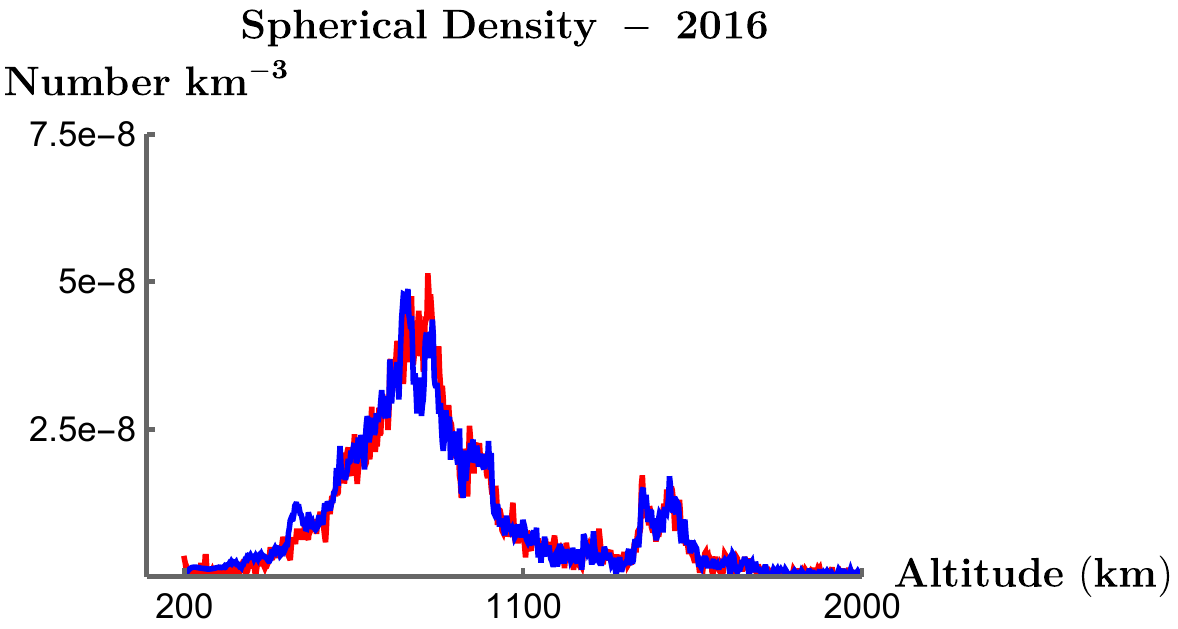}  }$
    $\hbox{\includegraphics[width=71mm]{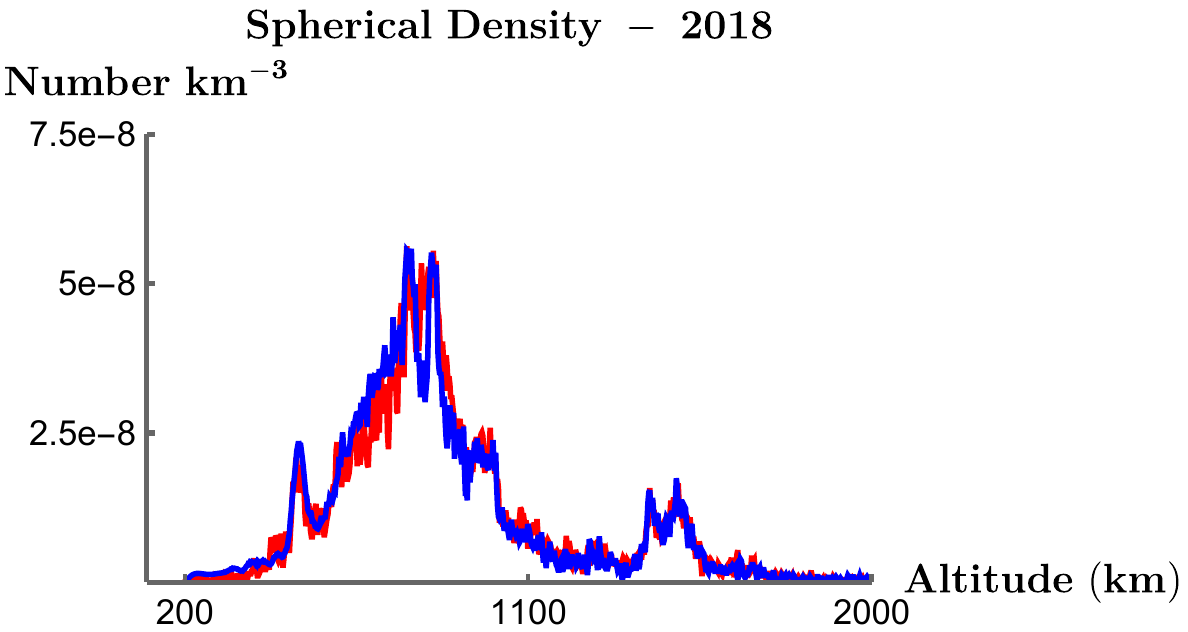}  }$
    $\hbox{\includegraphics[width=71mm]{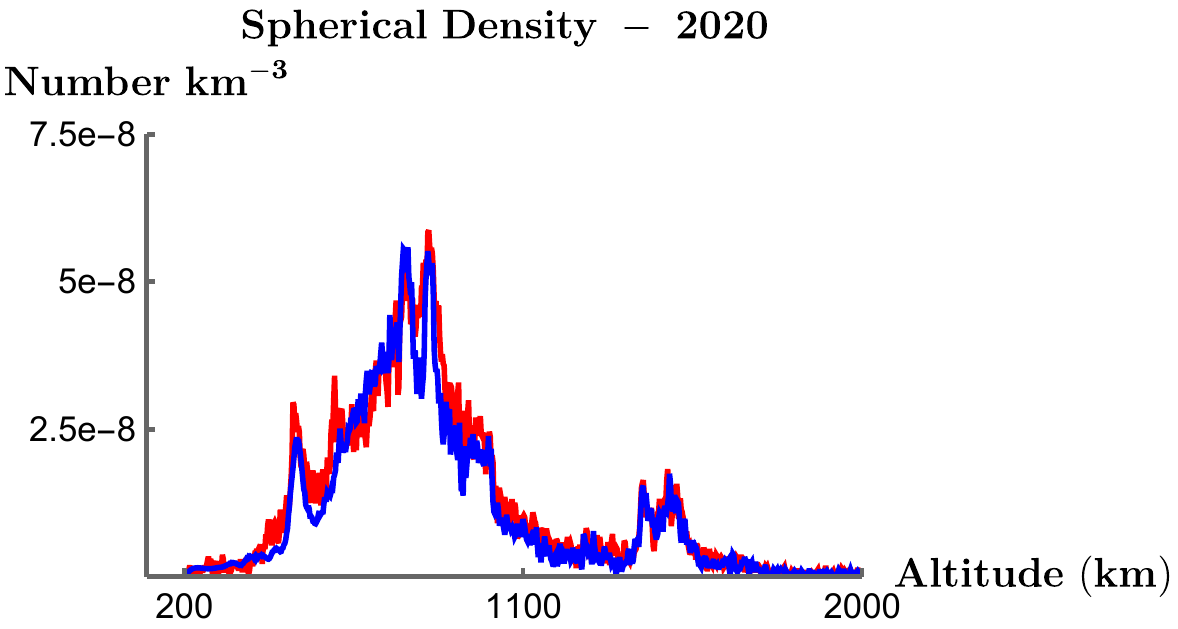}  }$
    $\hbox{\includegraphics[width=71mm]{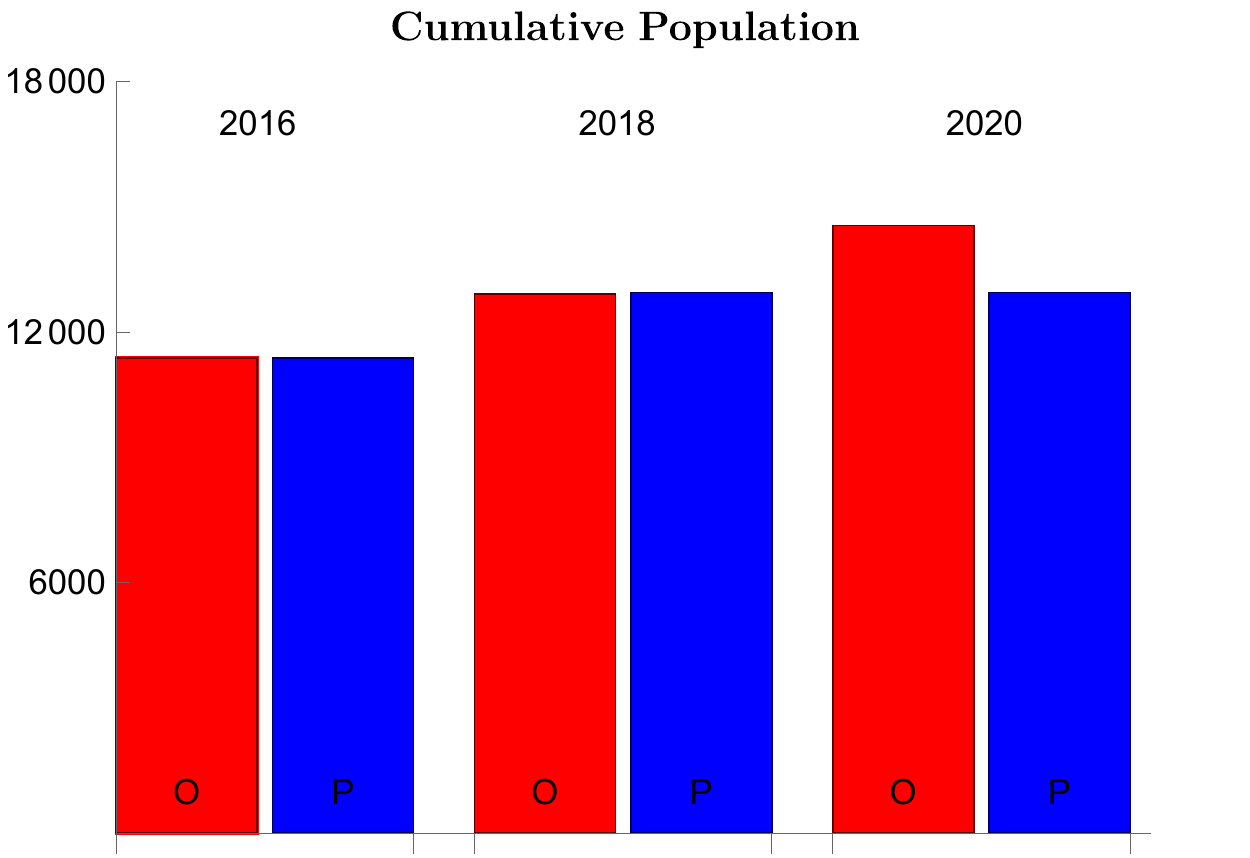}  }$
    \caption{ Comparison of observed (red) population and propagation of 2014 (blue) populations under the action of ground launches in 2016, 2018 and 2020 respectively, along with the  population growth (lower right panel). We observe an acceptable agreement between model predictions and the observations in both the $L^1$ and $L^\infty$-norms.
    }\label{Props}
\end{figure}

The red curves in Figure \ref{Props} show the evolution over the past 8 years of radial space debris density, as obtained from Space-Track.com \citep{SpaceTrack}. Note the steady, slow growth prior to 2022. The presence of the noticeable peak developing around 450 km at the 2022 epoch is due to the Russian test of their anti-satellite weapon November 2021, creating upwards of 1500 traceable pieces of debris in this altitude range \citep{Carnegie}. While such unitary and exceptional events are difficult to incorporate into a deterministic model, the model simulations show that these create differing initial conditions with drastically different possibilities for catastrophic growth. 

\begin{figure}[ht]
$\hbox{\includegraphics[width=71mm]{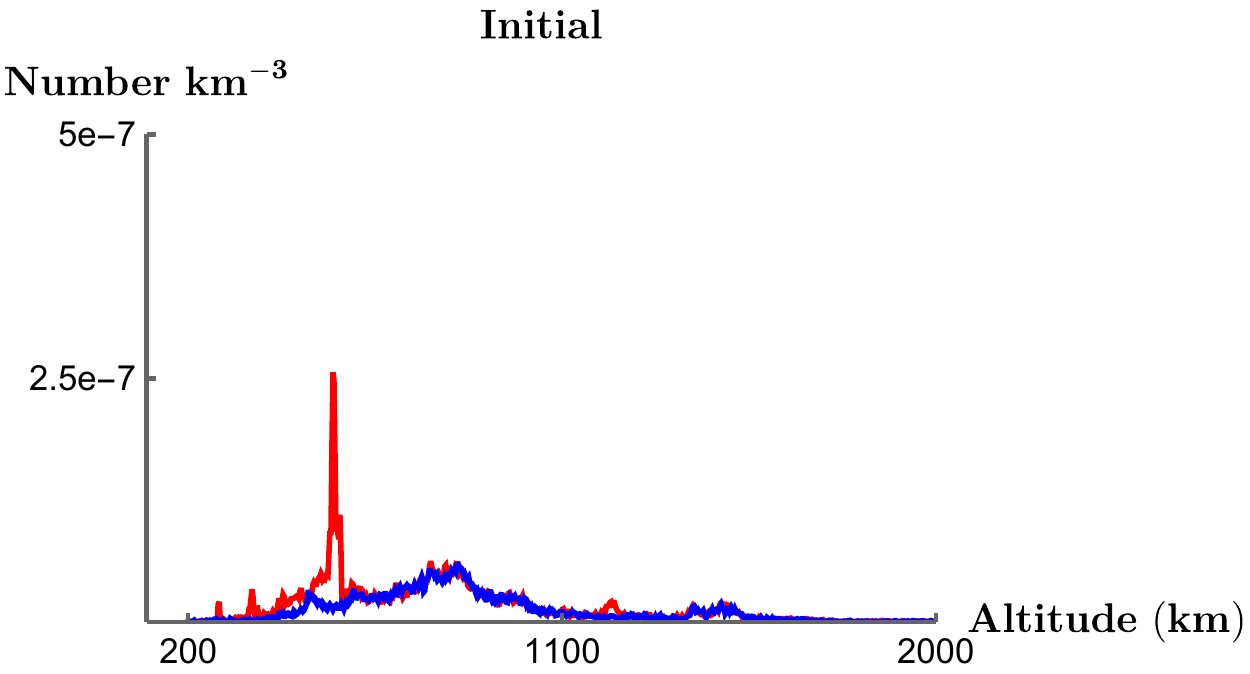}  }$
$\hbox{\includegraphics[width=71mm]{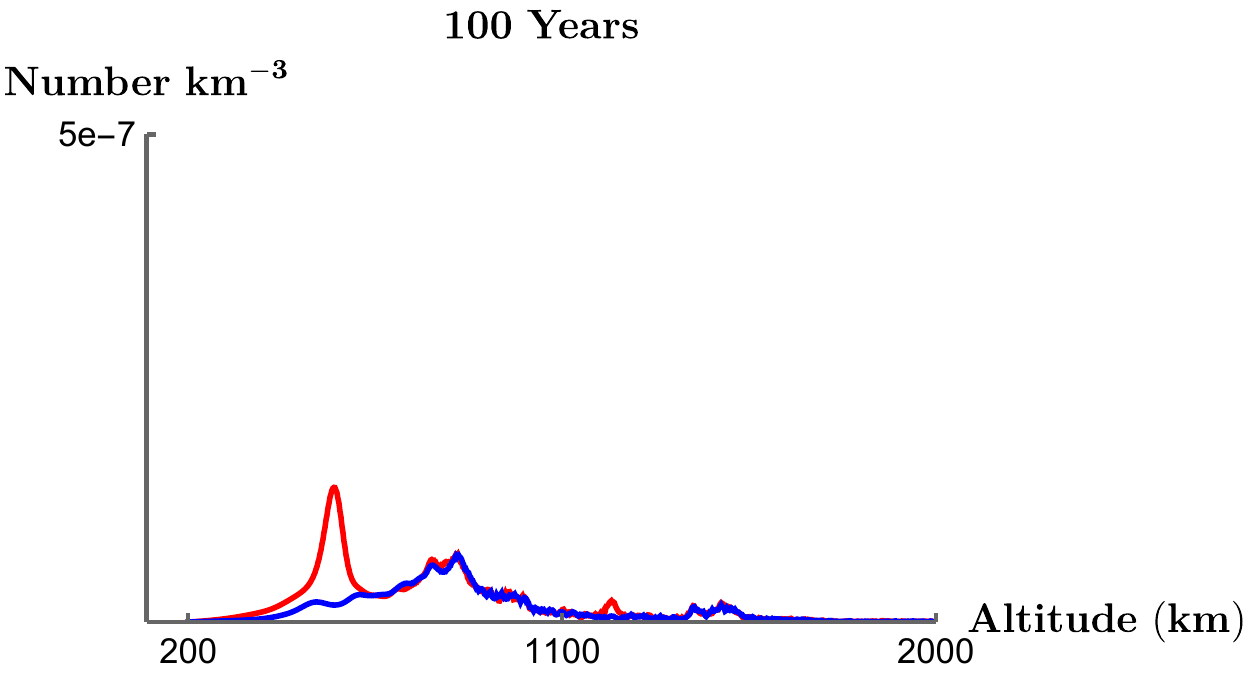}  }$\vspace{5mm}
$\hbox{\includegraphics[width=71mm]{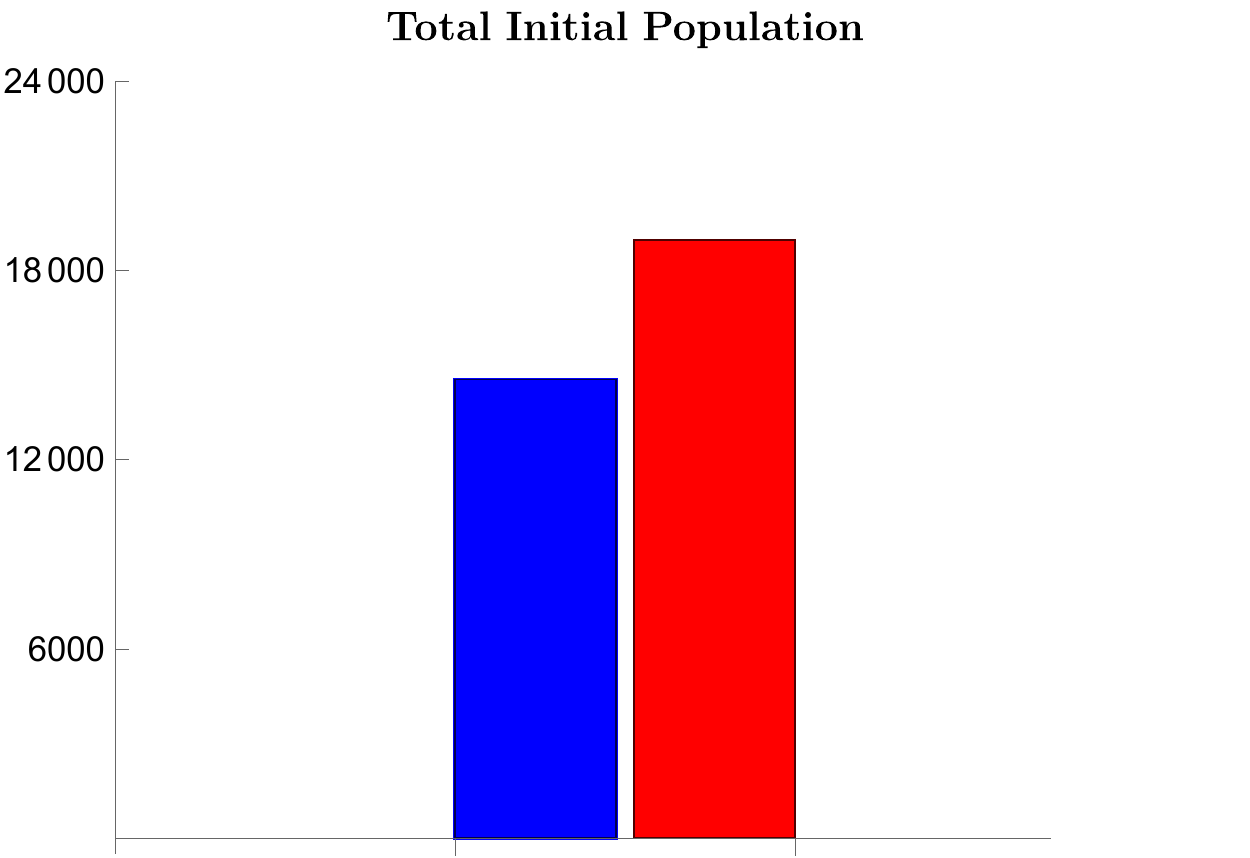}  }$
$\hbox{\includegraphics[width=71mm]{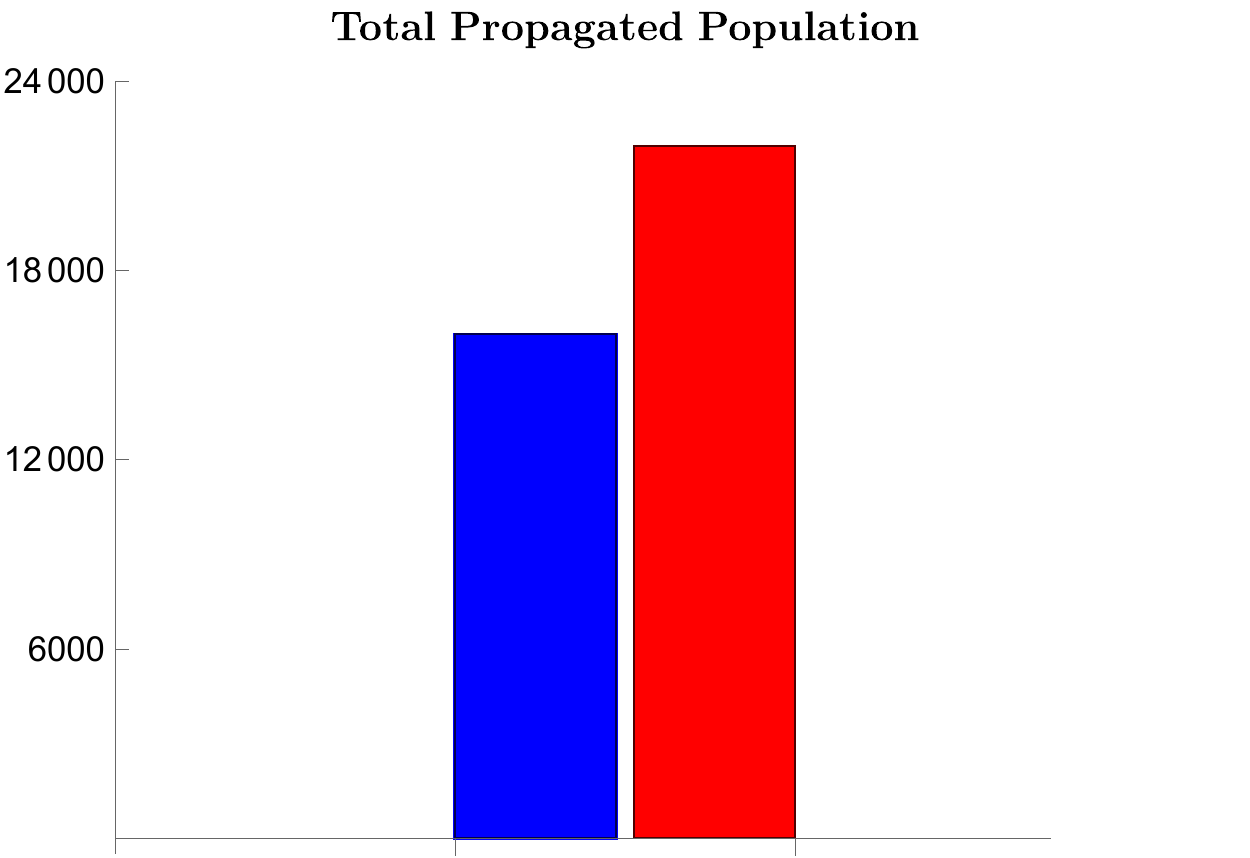}  }$

\caption{Initial densities (upper left) in 2020 (blue) and 2022 (red), along with densities propagated to 2120 (upper right) in the event of no further deposition The lower panels show the total populations for each scenario. Note that the 2022 growth is substantially greater due to the Kosmos-1408 debris spike at 500 km altitude. Note diffusion effects at the lower altitudes, and the general growth of the population as diffusion and collision occur. }\label{NoDepSims}
\end{figure}

\begin{figure}[ht]
    \centering
    \includegraphics[width=71mm]{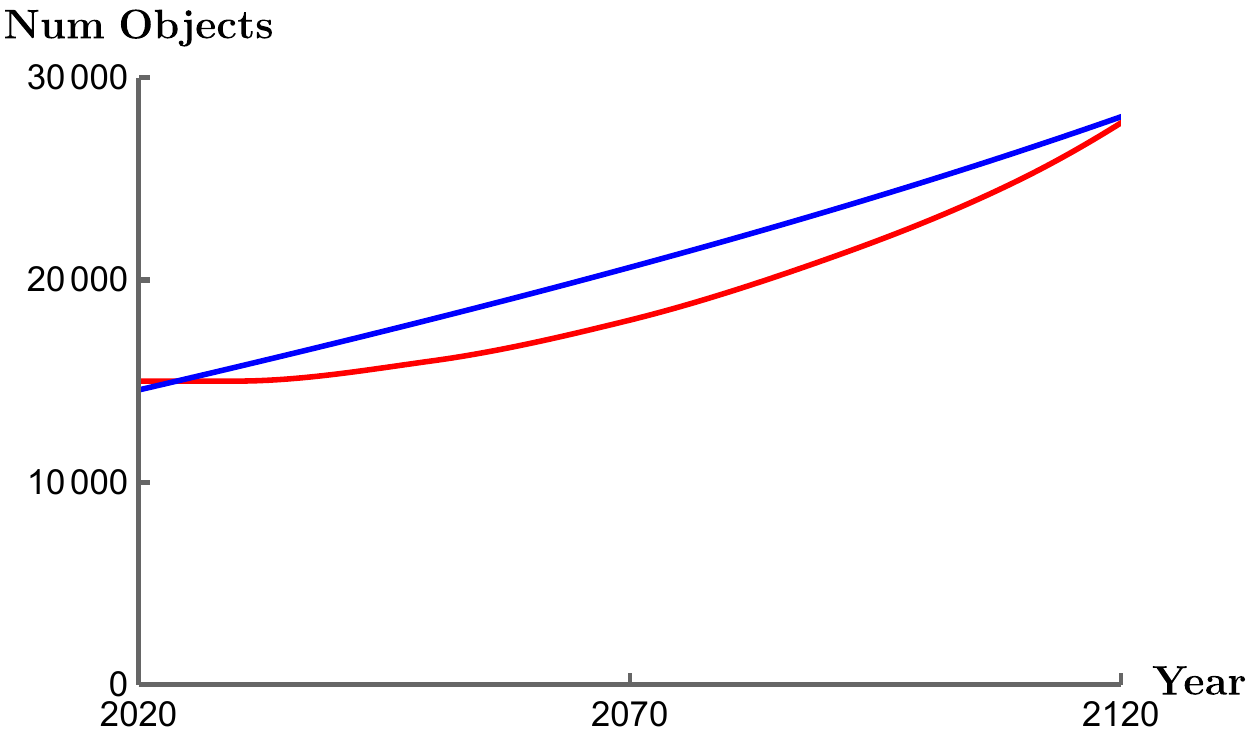}
    \caption{Comparison of total population forecasts for 2020 population by out model (blue) to the predictions of LEGEND (red) from 2020 to 2120.  The LEGEND data are reproduced from figures presented in \cite{Liou2011}. The launch rate for our simulations was intentionally reduced from the 2020 value for comparison to LEGEND.}
    \label{LEGEND}
\end{figure}

Figure \ref{LEGEND} compares the aggregate population projections of Equations \ref{mod1}-\ref{in_cond} to those of NASA's LEGEND model (which only reports the total population in a publicly) for the next 100 years. We observe  a close agreement between the predictions of the model versus those of LEGEND, deviating by a maximum of approximately 9\%. Note these simulations were prepared with the value of $T(t)$ artificially reduced to match the lower assumed launch rate at the time the LEGEND data were generated.

Figure \ref{NoDepSims} shows the evolution of the population starting from two different epochs, 2020 and 2022, respectively, in the absence of ground launches.  It is clear that the presence of the debris left from the Russian weapons test in 2021 causes significant hazard in the form of a persistent debris band, causing greater population growth through collisions.  

\begin{figure}[ht]
    \centering
    $\hbox{\includegraphics[width=71mm]{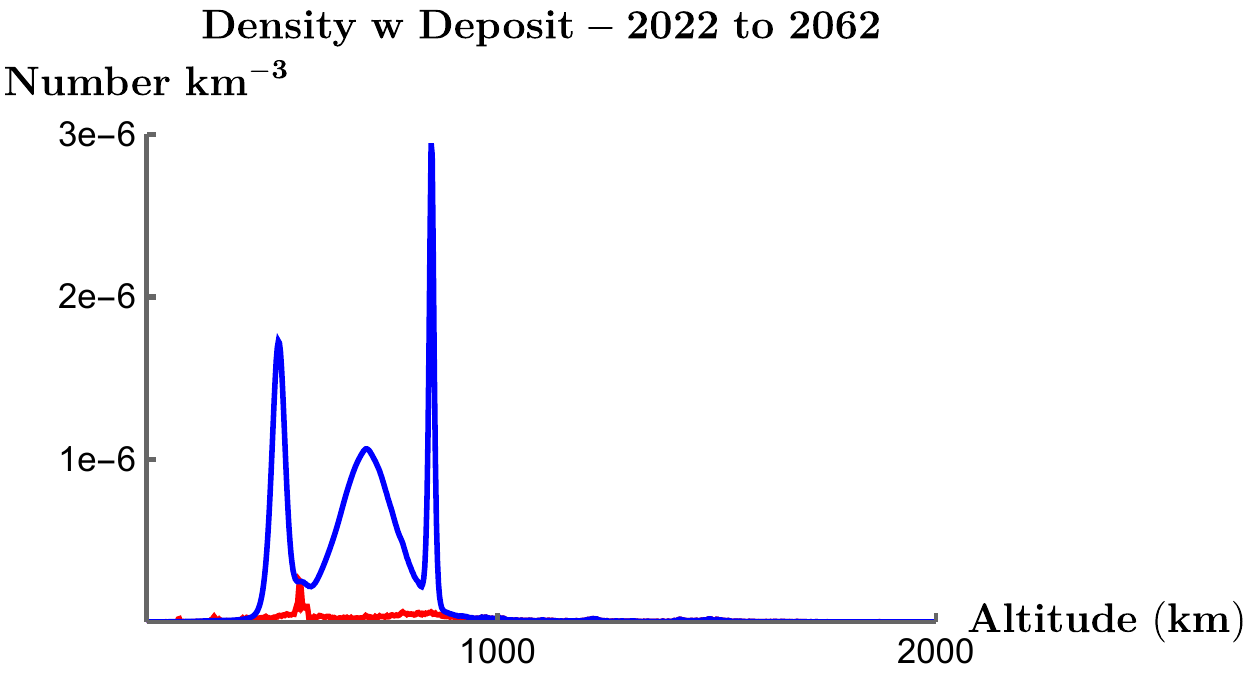}  }$
    $\hbox{\includegraphics[width=71mm]{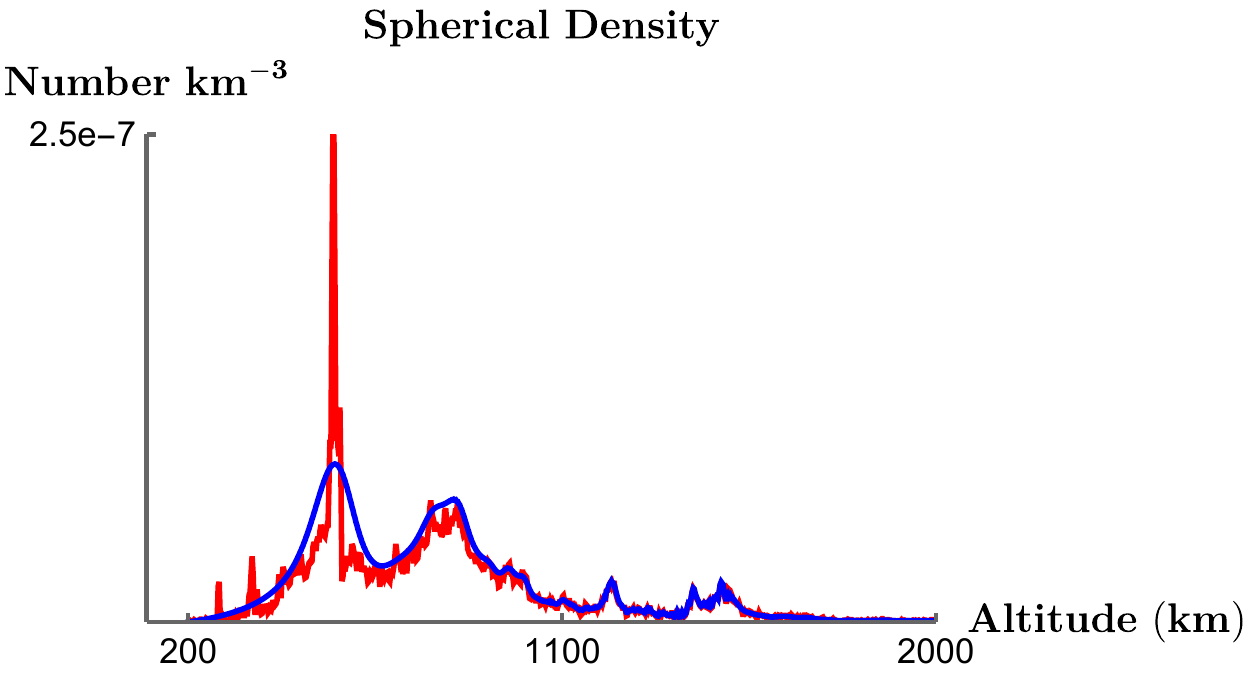}  }$
    \caption{\textbf{(Left)} Evolution of the 2022 populations (initial in red, propagated in blue) to 2062 under continued deposition at 2022 rate. The growth in this scenario over 40 years is a full order of magnitude greater than the no-launch scenario after 100 years, underscoring the danger of unmitigated launching. \textbf{(Right)} Propagation of the 2022 density to 2072 with $\alpha=0.878$ km$^2$ d$^{-1}$ and $\xi=10^{-3}$ km$^2$ d$^{-1}$, taking effect at $1500$ km. Compare this to Figure \ref{NoDepSims}, where the smaller diffusivity allows a greater buildup of population at lower altitudes. }\label{Growth_w_Dep}
\end{figure}

To analyze the effect of space policy on the debris population, we first assume that ground policy does not account for the removal of space debris. That is, $\Delta(r,t)$ is assumed to be always positive and of the separable form specified in Equation \eqref{deposfct}. Comparisons of the propagated and observed populations for 2016, 2018 and 2020 are shown in Figure \ref{Props}. Figure \ref{Growth_w_Dep} show the impact of continued deposition on the growth of the 2022 population. Catastrophic growth is triggered within the next 50 years according to our evolutionary model. 

\begin{figure}[ht]
\centering
$\hbox{\includegraphics[width=71mm]{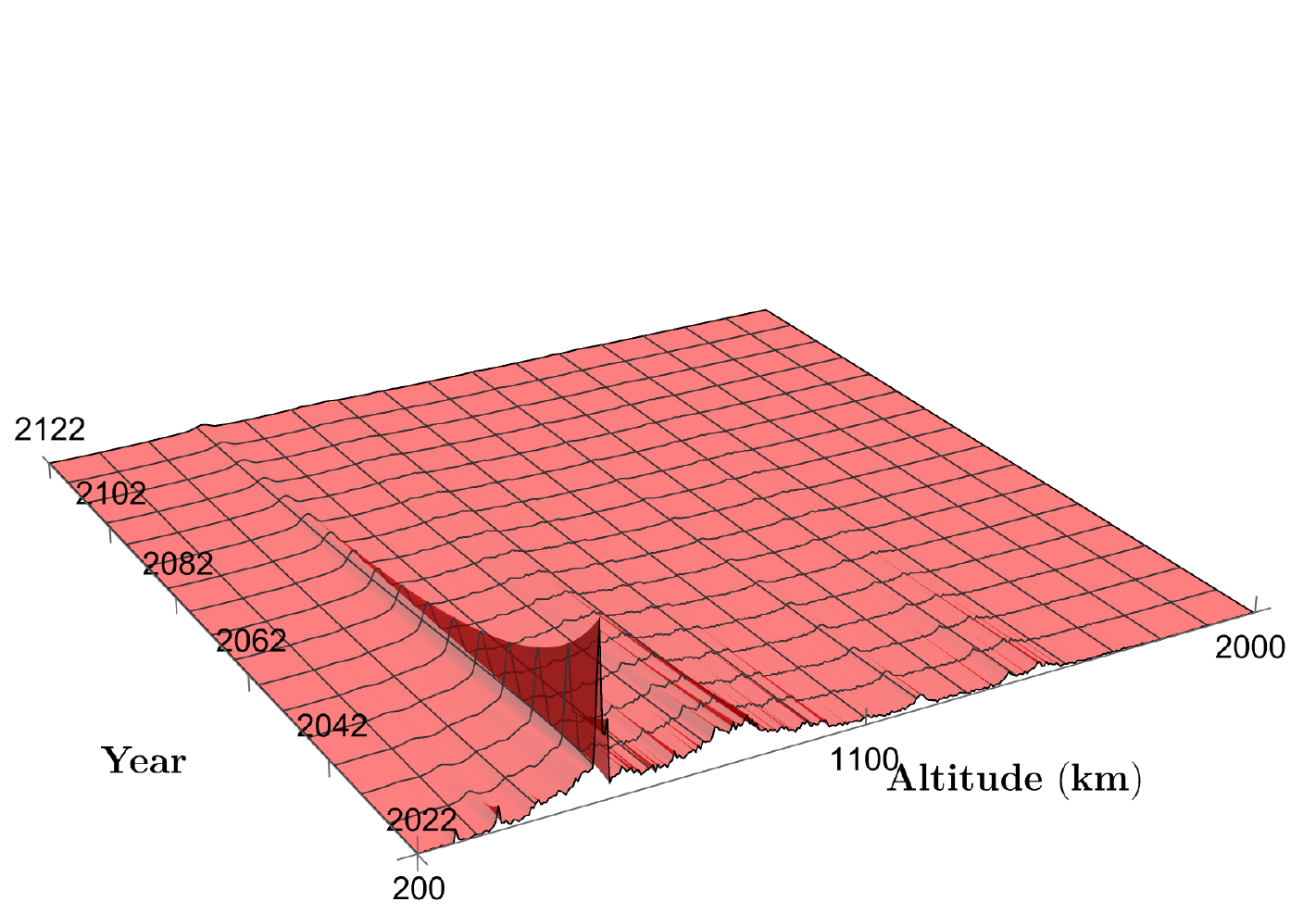}  }$
\caption{Propagation of the 2022 population to 2072 under action of active removal of the form $\Delta(r,t)=-\eta u(r,t)$. The  removal amounts to $5\%$ per year. Active removal strategies therefore curb the possibility for catastrophic growth.}\label{Remove}
\end{figure}

To investigate the effects of possible active removal strategies, we now allow $\Delta(r,t)$ to potentially take on negative values. As a first attempt, we assume the removal a certain fraction of the population, that is $\Delta(r,t)=-\eta u(r,t)$. Figure \ref{Remove} shows the effect of a 5\% per year overall removal on the population over the next 50 years. While a blanket removal strategy is not a feasible policy, this simulation shows that removal strategies are directly effective in the mitigation of catastrophic growth.   

\section{Discussion and conclusion}\label{Disc}

Model \eqref{mod1}-\eqref{in_cond} is a first step towards an accurate continuum model and is amenable to efficient numerical implementation. Further, the inclusion of the $\Delta(r,t)$ term allows for in-depth analysis of the effect of ground launch policy on future evolution. In particular, we see that continuing on the current course of launching will cause cascading growth to occur at a sooner epoch than if the debris population were left to evolve on its own. As only a few examples have demonstrated, an important part of future research is a careful determination of the diffusion parameter and, once the technology becomes widely available, the  debris removal rate.

Of particular interest to scientists and policymakers are strategies for reduction of debris input and the active removal of space debris. Such strategies are a subject of active research. \citet{Klima2018}, for example, approach the topic from a game theoretic perspective. As debris removal reduces the risk to all active satellites, every stakeholder may delay their action leading to a strategic dilemma. Especially due to the cost incurred in their use, such removal strategies, while physically possible as of writing, are still very much experimental. Still, despite these shortcomings, these new technologies represent the best feasible means to avoid the Kessler syndrome. A responsible stewardship of near-Earth space is clearly a task for public and private institutions alike that benefit from human space activity   \citep{Gast2022}.

The means of these removal strategies bears some mention. \citet{Bonnal2020} describe an orbiting laser station whose purpose is to improve knowledge about the trajectories of debris objects. There is also the possibility to impart velocity changes to objects using this orbital laser. \citet{Takahashi2018} propose a concept for space debris removal by a bi-directional plasma plume ejected from a satellite. One of the beams impacts the debris object while the other keeps the cleaning satellite in place. This has been demonstrated in a laboratory experiment.   All of these removal techniques act rather locally, as opposed to the global removal considered in this paper, but could be incorporated into further refinements of the model as time-varying velocity fields in addition to the passive removal of the air drag diffusivity currently present. 

There exist a number of directions for future refinement and expansion of the model.  The parametrization of the model may be fine-tuned as technology increases for detecting orbital objects, or as further data comes to light. The model simulations are sensitive to parameter values. Figure \ref{Growth_w_Dep} (right)  shows the effect of changing the parameters $\alpha$ and $\xi$, as well as the critical altitude in the diffusivity $D(r)$. Indeed, a greater diffusivity can arrest the growth due to collision events. 

%\begin{figure}[ht]
%\centering
%$\hbox{\includegraphics[width=71mm]{different_diffusivity.pdf}  }$
%\caption{Propagation of the 2022 density to 2072 with $\alpha=.878$ km$^2$ d$^{-1}$ and $\xi=10^{-3}$ km$^2$ d$^{-1}$, taking effect at $1500$ km. Compare this to Figure \ref{NoDepSims}, where the smaller diffusivity allows a greater buildup of population at lower altitudes. }\label{newDiff}
%\end{figure}

Further, Figure \ref{NoDepSims} demonstrates that, despite the high accuracy of the propagation model in accounting for collision events, certain events (such as unitary breakup or the Russian anti-satellite test) can occur with no advance warning and are not directly related to the population at any time. This hampers the ability to account for such events in a purely deterministic partial differential equation model. Therefore, the inclusion of stochastic impulses into $\Delta(r,t)$ could further enhance the accuracy and predictive power of the model. While the statistics of such events are at present not widely understood, this presents a very interesting and fresh area for investigation.  

Figure \ref{Remove} demonstrate the sensitivity of the debris population to the policy encoded by $\Delta(r,t)$. Because of the importance of policy to the overall evolution of the population, choosing an effective policy is a key strategic decision. Mathematically, we can formalize this choice into choosing a policy $\Delta$ that minimizes a functional
\begin{equation*}
I(\Delta)=\int_{t_0}^{t_1}\int_{r_E+200}^{r_E+2000}F(r,t,u,u_r,u_t,\Delta)r^2\,\ud r\,\ud t,
\end{equation*}
where the debris populations $u(r,t)$ is constrained by the model \eqref{mod1} - \eqref{in_cond}. By formulating a function $F$ that reflects some effective risk metric, the optimization problem described above can be used to directly inform policy-making.  

\section*{Author statement}
We declare that we have no competing interests.

\section*{Acknowledgments}
JJ is supported by a graduate fellowship from the Department of Mathematical Sciences at the University of Wisconsin - Milwaukee. We thank Drs.~Kerri Cahoy (Massachusetts Institute of Technology),  Colin McInnes (University of Glasgow) and  Jer-Chyi Liou (NASA) for sharing literature and valuable comments.  We also thank two unknown readers for constructive remarks.

\bibliography{literature}

\end{sloppypar}
\end{document}